\documentclass[preprint, superscriptaddress, showkeys]{revtex4}
\usepackage[pdftex]{graphicx}
\usepackage{amsmath}
\usepackage{amssymb}

\begin{document}

\title{Lymphotactin: how a protein can adopt two folds.}
\date{\today}

\author{Carlo Camilloni}
\email{cc536@cam.ac.uk}
\affiliation{Department of Chemistry, University of Cambridge, Lensfield Road, Cambridge, CB2 1EW, United Kingdom}
\author{Ludovico Sutto}
\email{ludovico.sutto@cnio.es}
\affiliation{Spanish National Cancer Research Center (CNIO), Structural Biology and Biocomputing Programme, Melchor Fernandez Almagro, 3. E-28029 Madrid, Spain}

\begin{abstract}

Metamorphic proteins like Lymphotactin are a notable exception of the empirical principle that structured natural proteins possess a unique three dimensional structure. In particular, the human chemokine lymphotactin 
protein (Ltn) exists in two distinct conformations (one monomeric and one dimeric) under physiological conditions. In this work we use a C$_\alpha$ G\=o model to show how this very peculiar behavior can be reproduced. From the study of the
thermodynamics and of the kinetics we characterize the interconversion mechanism. In particular, this takes place through the docking of the two chains living in a third monomeric, partially unfolded, state which shows a residual structure involving a set of local contacts common to the two native conformations.
The main feature of two--fold proteins appears to be the sharing of a common set of local contacts between the two distinct folds as confirmed by the study of two designed two--fold proteins.
Metamorphic proteins may be more common than expected.

\end{abstract}


\maketitle

\section{Introduction}

It has been recently shown the existence of few proteins with two stable native folds \cite{Murzin:2008p3201}. In particular, the human chemokine lymphotactin 
protein (Ltn) studied by Tuinstra et al. populates two well-defined conformations under physiological conditions \cite{Kuloglu:2001p10521,Kuloglu:2002p10522,Tuinstra:2008p1274}. One is a monomeric three-stranded $\beta$-sheet ending with an $\alpha$-helix at the C terminal (Ltn10). The second is a dimeric $\beta$-sheet conformation (Ltn40). The two structures are very dissimilar so as their biological function, which is to bind to glycosaminoglycans and to activate the Ltn XCR1 receptor respectively. 

The experimental observation of the presence of two distinct folds encoded in a single sequence, challenges the accepted vision of protein folding\cite{Anfinsen:1973p12079,Bryngelson:1987p12192}. In particular, are these proteins an exception to Anfinsen's paradigm or are they more common than expected? How can we explain this behavior in the current protein folding framework? We know that evolution moulds the rough energy landscape of a sequence leading a protein into a smoothed energy funnel removing most of the energetic contradictions\cite{Bryngelson:1995p12792,Leopold:1992p12793}. 
In the Ltn case, unlike in domain swapping proteins\cite{Bennett:1994p13128,Bennett:1995p13127}, the interactions that stabilize the homodimer are not borrowed from the single monomer chains. There rather exists two distinct sets of contacts for the monomeric and the dimeric conformations each stabilizing its respective structure.
What is then the strategy adopted by evolution to deal with the larger frustration of this 59-residues sequence still able to reversibly interconvert between two different functional folds?

In the present work, a computational model is used to shed light on these questions. The study of protein folding is classically based on the assumption that the native fold is unique and that this conformation is the only one minimizing the energetic contradictions. One simple implementation of this principle of minimum frustration is the G\=o model \cite{Go:1983p12078}. In such a model the potential is suggested by the native structure since only the native contacts contribute favorably to the energy of the system leading the native conformation to a global energy minimum. The G\=o model has been successfully used in different flavors to shed light on protein folding \cite{Clementi:2000p10423,Vendruscolo:2001p1976,Sutto:2006p255,Best:2008p304}, dimerization\cite{Yang:2004p12566} and aggregation \cite{Khare:2003p13206,Sharma:2008p13273}.

Here we show, for the specific case of the  Ltn10 - Ltn40 interconversion, that it is sufficient to merge the two sets of native contacts of the two distinct folds into a pool of equally weighted contacts in order to successfully reproduce the equilibrium population of the two folds.
The fact that such a naive extension of the G\=o model still works is not trivial since merging different sets of native contacts introduces frustration in the system. In general, not all the contacts can be satisfied by the same structure and moreover new energy minima may appear, leading to an ensemble of misfolded conformations. Nonetheless we observe a clear transition 
between the two folds in a precise range of temperature and monomers concentration. We reproduce the experimental situation where both native structures exist with equal probability and calculate the free energy of the system. On this surface, the denatured state displays peculiar features, quite different from those of two-state folders.
Analyzing this state both structurally and by means of a large number of kinetics runs connecting the two native folds, we suggest what is the mechanism of interconversion of Ltn10 monomers into Ltn40 homodimer.

To investigate if the possibility to host two native conformations is universal or is a property of specific proteins we designed two putative two-fold proteins. Observing that the interconversion Ltn10 - Ltn40 takes place through conformations characterized by a core of conserved common contacts we merged the contacts of Ltn40 with either the native contacts of src-SH3 or the contacts of Dendroaspin. Both proteins have the same length of a single Ltn chain but show different folds. We chose the src-SH3 because it partially shares the core of common contacts present in both Ltn folds while Dendroaspin has a completely different set of native contacts.
While in the first case this new protein still exhibits a two-fold behavior in the second case the system does not show any stable structure. The main feature of two--fold proteins appears to be the sharing of a common set of local contacts between the two folds.

These results not only clarify why merging the native contacts of two distinct fold worked in the Ltn10 - Ltn40 case but also point, more generally, to a simple strategy evolution can have adopted to confer proteins multiple folds and functions. Metamorphic proteins may not be as rare as it currently seems\cite{Murzin:2008p3201}.

\section{Materials and Methods}

Lymphotactin shows two distinct native conformations. In fact, two 59 residues monomers (Ltn10) made of three $\beta$--strands and an $\alpha$--helix can interconvert into a homodimer (Ltn40) where each monomer is a four strands $\beta$--sheet. Notably, the two states are stabilized by a different set of native contacts. In our model Ltn10 shows 146 native contacts per monomer while the dimer fold Ltn40 has 372 native contacts coming from 125 tertiary contacts within each monomer and 122 quaternary contacts corresponding to the interface. Only 38 tertiary contacts are common between the two folds. These interactions involve the disulphide bond between Cys 11 and Cys 48 and a set of contacts between $\beta$1 and $\beta$3. Moreover the homodimer exhibits a set of quaternary native contacts \cite{Tuinstra:2008p1274}. As a consequence, Ltn40 is energetically favored with respect to Ltn10 while being entropically discouraged.
Having defined these two distinct sets of native contacts, we introduced a G\=o potential given by the union of the two sets~\cite{Zuckerman:2004p12565} that can be formally written as:
\begin{equation}
H_{Go}=-\sum_{pairs}^{Ltn10 (1)}\epsilon - \sum_{pairs}^{Ltn10 (2)}\epsilon - \sum_{pairs}^{Ltn40}\epsilon + angles + dihedrals ,
\end{equation}
where the contacts in common between the monomers Ltn10 and the dimer Ltn40 are counted only once and where each contact has the same weight $\epsilon$. We chose a C$\alpha$ coarse grained description of the protein to keep the description as simple as possible.

The simulations are performed using the GROMACS package \cite{Hess:2008p191} merging the topologies of the two proteins (pdb code: 2HDM, 2JP1)\cite{Kuloglu:2001p10521,Tuinstra:2008p1274} resulting from the Onuchic' structure based potential software \cite{Clementi:2000p10423,Whitford:2009p11363}. The distance between two consecutive C$\alpha$ was kept fixed using the LINCS algorithm \cite{Hess:1997p5391} while the disulphide bond between Cys 11 and Cys 48 of both monomers is implemented with a restraining potential that is zero when the respective C$\alpha$ are within the distance range $[r_{Ltn40};r_{Ltn10}]$ and that is harmonic outside this interval. 

We performed two parallel tempering \cite{Hansmann:1997p7019} Langevin dynamic simulations of $2\cdot10^8$ steps with a time-step of 2 fs using eight different temperatures (120, 127, 134, 142, 150, 159, 168, 178 K); $T_c=134$K is chosen as the reference temperature for the system and in what follows the temperatures are expressed in units of $T_c$. The two parallel tempering were performed at two different concentrations (i.e. 0.46mM and 0.06mM). The effective concentration is obtained by restraining the distance $d_{CM}$ between the centers of mass of the two monomers with an harmonic potential for $d_{CM}>$12 nm and $d_{CM}>$24 nm respectively. 

We simulated 100 trajectories, starting from two monomers, which were stopped once the protein finds its dimeric fold, for a total of $10^9$ steps. These simulations were done at $T=1$ and with a concentration of 0.06mM.

To study to what extent the two--fold behavior of Lymphotactin proceeds from its particular native structure we generated two putative two--fold proteins by merging the native contacts of the dimer Ltn40 conformation with either the src-SH3 (pdb code 1FMK) or Dendroaspin (pdb code 1DRS). These two new proteins were simulated at a concentration of 0.46mM for two different temperatures $T=1$ and $T=1.04$ and for $10^9$ steps each.

In order to study the interconversion between the two known folds of Lymphotactin we reconstruct the free energy surface (FES) along the following set of reaction coordinates. The two RMSD with respect to the native structure of Ltn10 (called RMSD-monomer) and the two RMSD with respect to each single chain of Ltn40 (RMSD-dimer) allow to distinguish the two tertiary conformations for the two chains. 
The ratio $r_{mM}=d_m/d_M$ of the minimum and maximum distance between the two chains is then chosen as a fifth reaction coordinate in order to study the docking between the two monomers.

To compare the different structures on a coarse-grained contact level, we also use the fraction of formed native contacts $q$ grouped in four disjoint sets. In particular, $q_{monomer}$ is calculated counting the 108 native contacts specific to the Ltn10 monomeric conformation, similarly $q_{dimer}$ counts the fraction of the 87 native contacts specific to a single chain in the dimeric Ltn40 conformation, $q_{common}$ groups the 38 common native contacts between Ltn10 and Ltn40 structures and $q_{iface}$ accounts for the 122 quaternary native contacts of the dimeric interface. We indicate with $Q_x$ the average of $q_x$ over a set of conformations.

\section{Results and discussion}

\begin{figure}[htbp]
\begin{center}
\includegraphics*[width=\textwidth]{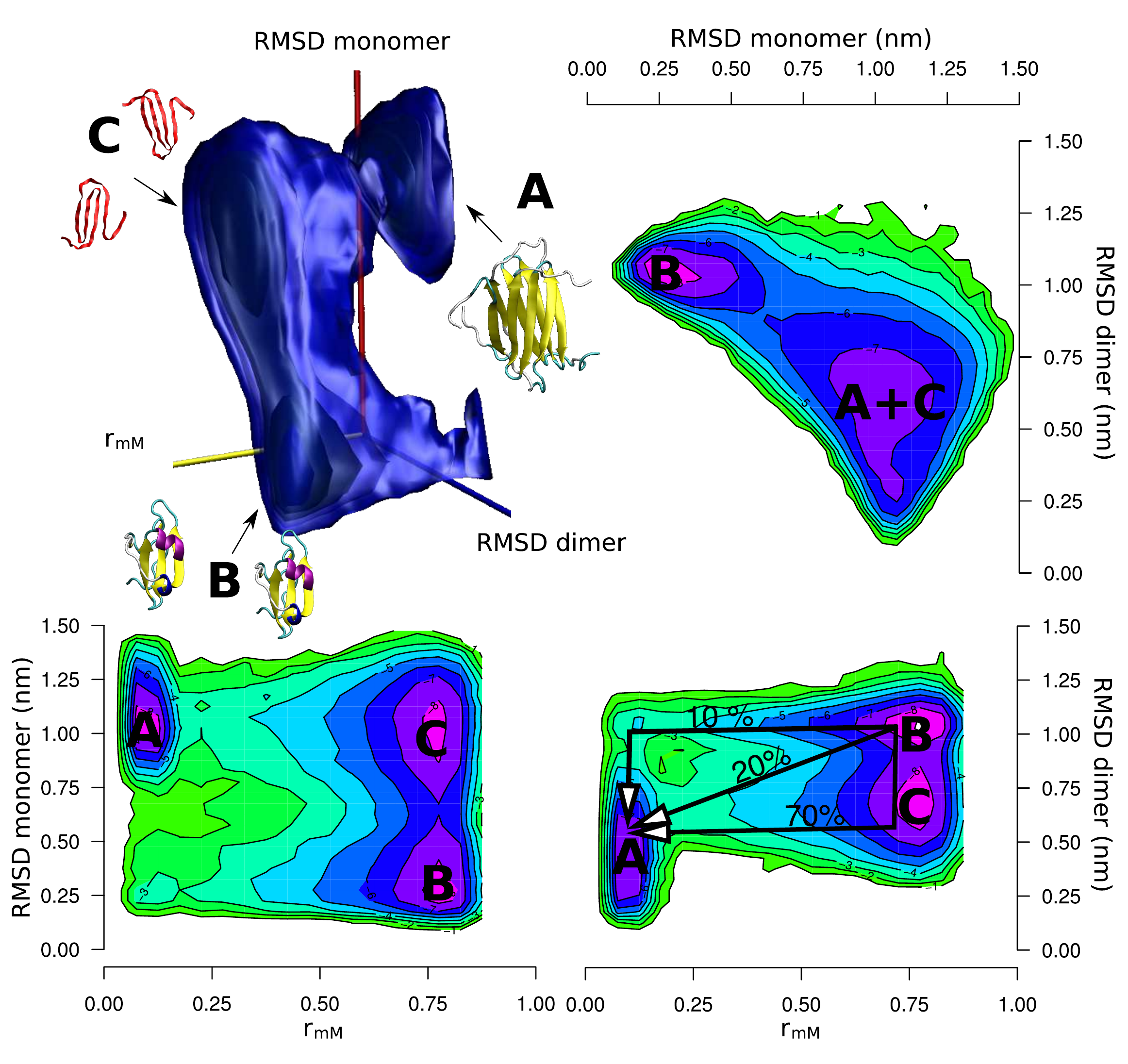}
\caption{The free energy of the two chains as a function of the three different order parameters at T=1 (0.06mM). The tridimensional image in the upper left corner captures the three equilibrium states of the system depicted with a cartoon representation aside. The state A corresponds to the homodimer Ltn40 fold, state B to the ensemble of conformations where at least one chain is in the Ltn10 fold and state C to an ensemble of monomeric conformations where at least one chain is unfolded. The bidimensional plots are the projections of the free energy along two of the three order parameters allowing for a more quantitative analysis. The free energy isolines are drawn every kT in all graphs. In the bottom right FES are also shown the pathways and their relative weight connecting B to A.}
\label{fig:1}
\end{center}
\end{figure}

At $T=1$ and for a concentration of 0.06mM the model resembles the biological condition in which the two Lymphotactin folds Ltn10 and Ltn40 are sensibly populated. In order to show the free energy surface (FES) we integrate over RMSD-monomer and RMSD-dimer relative to the second chain. In \ref{fig:1} is shown the resulting FES as a function of the three remaining reaction coordinates. We name A, B and C the  three well--separated minima observed in the figure. The one at low RMSD-dimer and low $r_{mM}$ (state A) corresponds to the dimeric Ltn40 fold while the one with low RMSD-monomer and high $r_{mM}$ is the monomeric Ltn10 fold (state B). A third state (C) corresponds to the two undocked chains (high  $r_{mM}$), where each chain is structurally different from the single chain of Ltn40 (RMSD-dimer > 0.6 nm) and where at least one chain is structurally different from Ltn10 (RMSD-monomer > 1 nm). At this temperature the A, B and C states are populated at 20\%, 40\% and 40\% respectively.  With respect to states A and B, the monomeric state C is structurally more heterogeneous. In fact, the average RMSD calculated between all pairs of conformations in state C is 0.7nm as compared to 0.3nm and 0.4nm for states A and B respectively (see \ref{tab:1}). Increasing the temperature, the free energy minimum of state C spreads towards higher values of RMSD-monomer and RMSD-dimer while the free energy basins of A and B shrink.

The reconstructed FES suggests that the Ltn40 dimer is formed by the docking of two chains in state C.
In order to assess the interconversion pathway we performed 100 more simulations starting from 2 monomers in the Ltn10 conformation (state B) and stopping them upon reaching the folded dimeric Ltn40 conformation (state A) (cf. Material and Methods).
We show in \ref{fig:1} that in the 90\% of the trajectories the Ltn40 fold is eventually reached through the transition of the barrier  located at $r_{mM}=0.25$ and RMSD--dimer=0.6nm. This corresponds to the docking of two chains structurally similar to C. In particular, in the 70\% of these trajectories the monomer interconverted from B to C before the dimerization while in the 20\% of the cases the monomer crosses the free energy barrier coming directly from the B state. The remaining 10\% of the trajectories reaches the Ltn40 conformation through a second transition state situated at  $r_{mM}=0.2$ and RMSD--dimer=1.0nm corresponding to the docking of two chains structurally similar to B.
These results underline the importance of state C in the dimerization process of the Ltn protein being an en-route ensemble connecting the monomer to the dimer state. 

\begin{figure}[htbp]
\begin{center}
\includegraphics*[width=\textwidth]{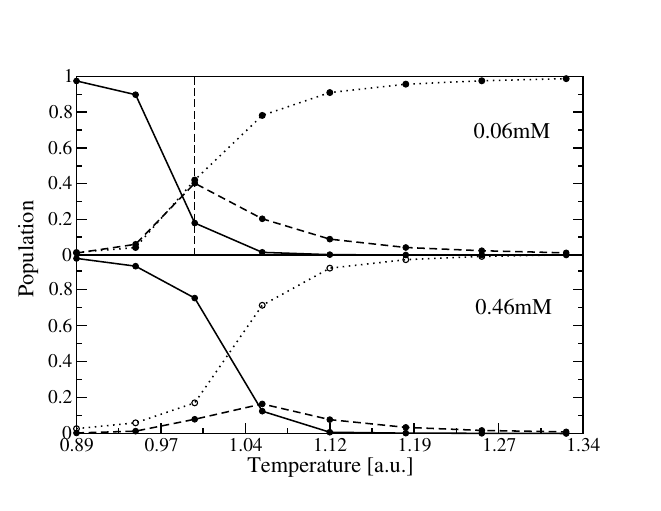}
\caption{Normalized population of the three states as a function of temperature in units of $T_c$ for two different concentrations. The A state (Ltn40) is indicated with a solid line, the B state (Ltn10) is shown with a dashed line and the C state is displayed with a dotted line. The vertical dashed line at T=1 in the low concentration panel (upper panel) indicates the conditions of the studied FES.}
\label{fig:2}
\end{center}
\end{figure}

In \ref{fig:2} are displayed the populations of the three states as a function of temperature for two different concentrations.
The dimer is preponderant over the monomer at low temperature. Increasing the temperature there is a maximum in the monomeric (Ltn10) population while the third state C increases monotonically becoming the only populated state at high temperature. In the case of lower concentration (\ref{fig:2}, upper panel) there is a temperature (i.e. 0.98) in which the three states are equally populated. 
The dimeric fold (Ltn40) has the lowest potential energy (the dimer has more than two times the number of contacts of the monomer) thus resulting favored by an increase of the concentration. 
Experimentally it is known that the Ltn10 is the most populated fold at $10\,^{\circ}\mathrm{C}$ and 200mM NaCl while at higher temperature ($25-37\,^{\circ}\mathrm{C}$) and in absence of salt the dimer Ltn40 is preponderant \cite{Tuinstra:2008p1274}. The fact that a dimer conformation is preponderant at high temperature where it is entropically discouraged with respect to two monomeric conformations cannot be understood without the introduction of entropy dependent forces like hydrophobicity. Moreover to include the effects of charged ions one should add an electrostatic potential. As a consequence it is not unexpected that our model designed to focus on the topological and energetic factors, cannot reproduce this behavior.

So far the C state can be interpreted as the denatured state from which the protein can either fall in the dimeric or the monomeric state. The presence of residual structure in the C state could help us shed light on the underlying mechanism of the interconversion. 
Adopting the Ltn10 and Ltn40 conformations as references, we defined the average fraction of native contacts $Q$ present in a set of conformations using four disjoint sets of native contacts. The grouping is done according to the specific presence of a native contact in only the Ltn10 fold ($Q_{monomer}$), only the Ltn40 conformation ($Q_{dimer}$) or in both structures ($Q_{common}$). A fourth parameter accounts for the fraction of native contacts of the dimeric interface $Q_{iface}$ (see Methods). The results are reported in \ref{tab:1}.
Interestingly, the third state C appears to be a monomeric state sharing a small number of native contacts specific to the Ltn10 conformation (low $Q_{monomer}$) while showing a $Q_{common}$ similar to both the conformations of states A and B and a sensible preference towards the tertiary structure of the Ltn40 fold. In fact half of the specific native contacts of the dimeric conformation are already present in the C state. In general, all three states keep the core of common native contacts mostly formed ($Q_{common}\approx0.7)$.

\begin{table}[htdp]
\begin{center}
\begin{tabular}{|c|c|c|c|c|c|c|}
\hline
State & $Q_{monomer}$ & $Q_{common}$ & $Q_{dimer}$ & $Q_{iface}$ & RMSD & N \\
\hline
A & $0.20\pm0.05$ &  $0.69\pm0.08$ & $0.72\pm0.08$ &  $0.77\pm0.09$ & $0.4\pm0.2$ & 1968 \\
B & $0.71\pm0.06$ &  $0.73\pm0.07$ & $0.19\pm0.03$ &  $0.00$ & $0.3\pm0.2$ & 5980 \\
C & $0.30\pm0.06$ &  $0.69\pm0.08$ & $0.50\pm0.09$ &  $0.00$ & $0.7\pm0.3$ & 5963 \\
\hline

\end{tabular}
\end{center}
\caption{Characterization of the three states. 
$Q_{monomer}$ and $Q_{dimer}$ are the average fraction of native contacts specific for the tertiary structure of the monomer and the dimer respectively. $Q_{common}$ is the average fraction of native contacts that are in common with the Ltn10 and Ltn40 conformations and $Q_{iface}$ is relative to the quaternary contacts between the two chains of the dimeric Ltn40. The RMSD column is the average internal RMSD of the set in nm and N the number of conformations of each set.}
\label{tab:1}
\end{table}

\begin{figure}[htbp]
\begin{center}
\includegraphics*[width=\textwidth]{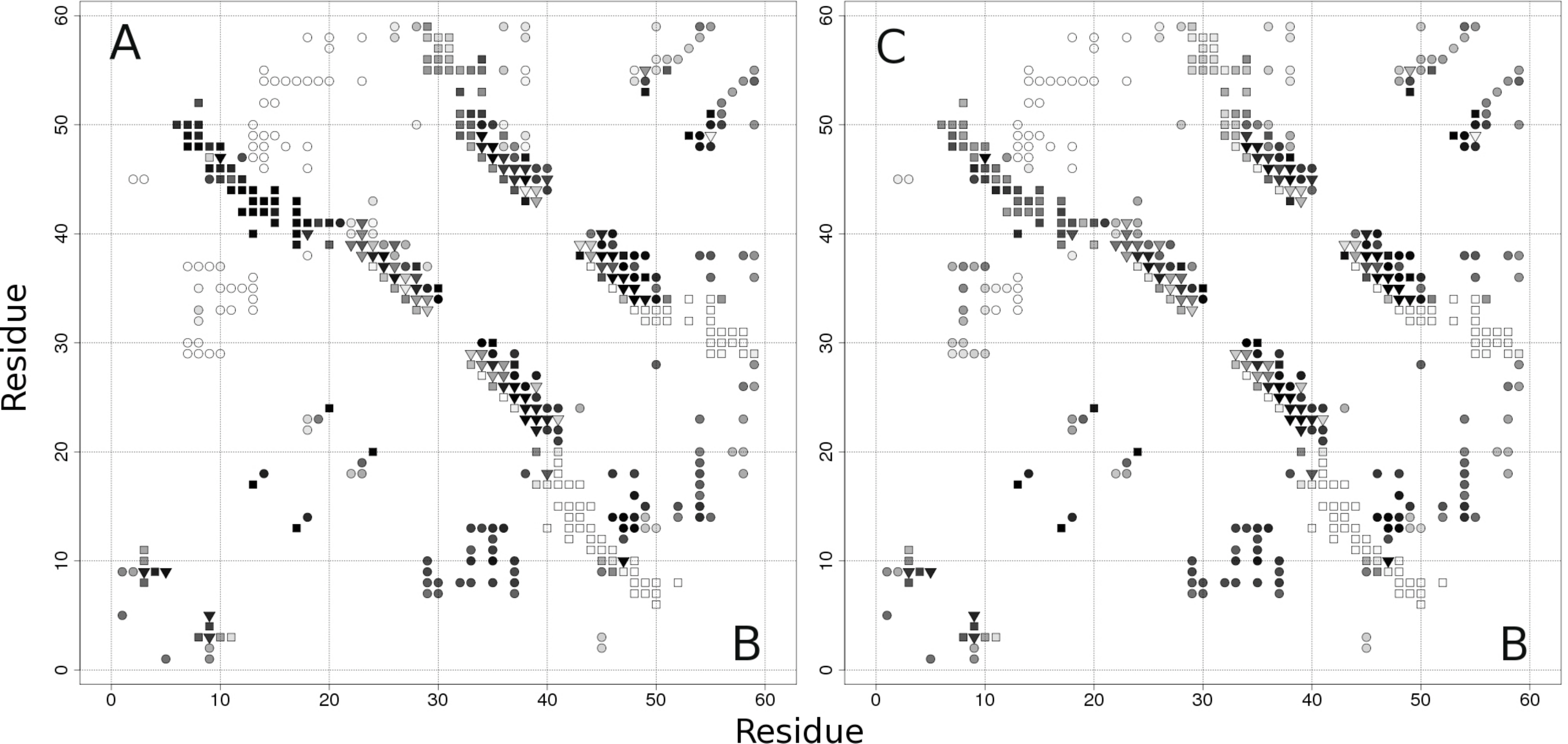}
\caption{Contact maps of states A, B and C. In the maps the squares represent the contacts belonging to the Ltn40 fold (state A), the circles are the contacts belonging to Ltn10 fold (state B) while the triangles are the contacts in common between the two folds. The probability of formation (i.e. stability) of each contact in the two states is represented by a gray scale going from the most rarely formed (white) to the contacts always present (black).}
\label{fig:3}
\end{center}
\end{figure}

The analysis of the single contact formation is reported in \ref{fig:3} with the contact map representation of states A, B and C. All three states share a central $\beta$--sheet that involves the common contacts (triangles in the figure) plus a set of specific contacts of either Ltn10 (circles) or Ltn40 (squares). In the case of state B the contacts within the central $\beta$--strand are shifted by one residue with respect to state A. 
Furthermore, the A and B states are mainly characterized by the presence of their peculiar set of tertiary structure contacts. In fact, the formation of local contacts is less specific since each state has both its own local contacts and contacts of the others states. From the second panel of  \ref{fig:3}, we observe that the stability of the local contacts of the C state is very similar to the previous cases. The C states conserves a core of local common contacts while the tertiary contacts have a preference toward the A state but are less stable (more filled squares than circles but lighter than in the A state).

\begin{figure}[htbp]
\begin{center}
\includegraphics*[width=\textwidth]{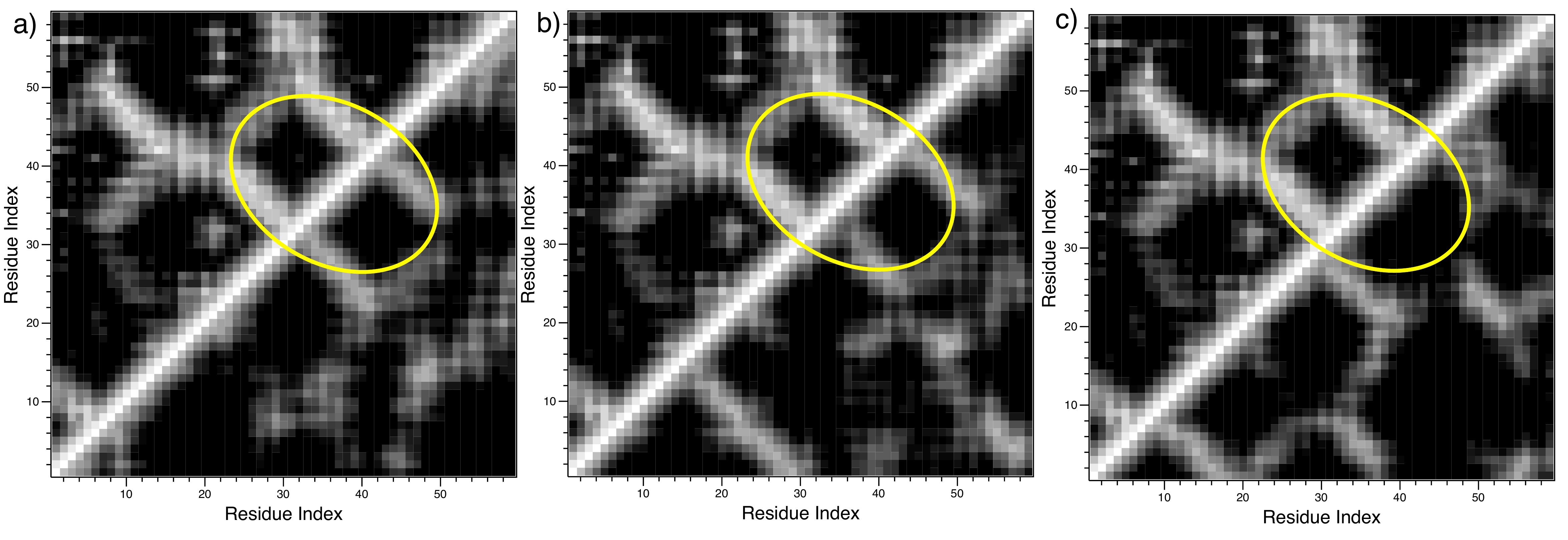}
\caption{Contact maps of Ltn40 vs Ltn10 (a), Ltn40 vs src-SH3 (b) and Ltn40 vs Dendroaspin (c). The circled region highlights the presence (in a and b) or the absence (in c) of the nucleus of local contacts. The gray scale measures the distance between residues.}
\label{fig:4}
\end{center}
\end{figure}

This analysis suggests us how evolution coped with the energetic frustration due to the presence of the two encoded folds in the Ltn sequence. On one hand a core of local contacts is conserved among the two folds reducing the accessible conformational space. On the other hand the non-local contacts are as different as possible giving rise to mutually exclusive secondary, tertiary and eventually quaternary structure leading to dimerization.
This scheme would suggest that mixing the native contacts of two distinct folds sharing only a common set of local contacts would lead to the same results. On the contrary, merging the contacts of two proteins without this constraint could lead to a highly frustrated system without any defined stable structure. To support this thesis we designed two putative two-fold proteins. We merged the contacts of Ltn40 with either the native contacts of src-SH3 or the contacts of Dendroaspin. These two latter proteins have the same length of a single Ltn chain while exhibiting a different native structure. We chose the src-SH3 because it partially shares the core of common contacts present in both Ltn folds as it clearly emerge comparing the contact maps shown in \ref{fig:4}(a) and (b). On the contrary, Dendroaspin has a completely different set of native contacts as can be seen from \ref{fig:4}(c).
To be noted that the interconversion from Ltn40 to both src-SH3 and Dendroaspin would be prevented by the presence of the disulphide bond between Cys 11 and Cys 48 since these residues are far apart in these two proteins. For this reason, the bond was removed without relevant consequences for the Ltn interconversion mechanism (data not shown).
The two new proteins were simulated at both T=1 and 1.04 and at a concentration of 0.64mM.

The Ltn40 - SH3 protein keeps half of the common contacts of Ltn40 - Ltn10 while substitutes the 108 specific contacts of Ltn10 with 150 src-SH3 specific contacts. Interestingly this new protein still exhibits a two--fold behavior, with a dimeric Ltn40 free energy minimum, a monomeric src-SH3 state and a third state in between.
In the second case, the Ltn40 - DRS protein has only 7 common contacts with the Ltn40 - Ltn10 system while the Dendroaspin conformation is characterized by 147 specific contacts. In this case the protein does not show any stable structure neither at the low nor at the high temperature. The presence of a set of local contacts conserved between two distinct folds seems to be the ingredient to build a two--fold protein.

The Ltn interconversion example can be viewed as a double basins system as in the case of proteins with large scale conformational changes but where the transition adds a new biological function. The latter systems have already been studied in a multiple basin energy landscape perspective \cite{Best:2005p12413,Okazaki:2006p12405}. 
From our study it emerges that two different folds can be easily accommodated in a single sequence provided some constraints are met. As a consequence, our model suggest that if these metamorphic proteins are rare it is not for the complexity of merging different folds on a sequence but the answer is probably to be found in an evolutionary basis.

\section{Conclusions}

Lymphotactin exists in equilibrium in two distinct native structures. By means of a structurally based model we have shown that at the core of the interconversion mechanism lies a set of conserved local contacts. In fact, the switching between the monomeric and the dimeric conformations takes place through a third partially unfolded state in which only the common contacts are stably formed. These metastable conformations find each others and dock leading to the folded Ltn40 homodimer after rearrangements. The two-fold feature can be reproduced by a designed protein as long as a core of local contacts is conserved.
This suggests a simple strategy evolution can have adopted to confer natural proteins multiple folds and functions.

\section{acknowledgement}

The authors thank G. Tiana and F.L. Gervasio for useful discussions and acknowledge the computer resources provided by the Centro de Supercomputaci\'on y Visualizaci\'on de Madrid (CeSViMa) and the Spanish Supercomputing Network as well as the Consorzio Interuniversitario Lombardo per l'Elaborazione Automatica (CILEA). C.C. is supported by a FEBS long--term fellowship.

\end{document}